\documentclass{article}
\usepackage{ijcai22}

\usepackage{times}

\usepackage{soul}
\usepackage{url}
\usepackage[hidelinks]{hyperref}
\usepackage[utf8]{inputenc}
\usepackage[small]{caption}
\usepackage{graphicx}
\usepackage{amsmath}
\usepackage{booktabs}
\usepackage{dirtytalk}
\usepackage{balance}

\usepackage{wrapfig}

\usepackage{caption}
\usepackage[separate-uncertainty=true]{siunitx}
\usepackage{mathtools}
\usepackage{graphicx}
\usepackage{courier}
\usepackage{epstopdf}
\usepackage{color}
\usepackage{csquotes}

\usepackage{amsmath}
\usepackage{amsfonts}
\usepackage{amssymb}
\usepackage[subnum]{cases}
\usepackage{booktabs} 
\usepackage{pifont}

\usepackage{algorithm,algorithmicx,algpseudocode}

\usepackage{amssymb}
\usepackage{pifont}

\usepackage {tikz}
\usetikzlibrary {positioning}
\definecolor {processblue}{cmyk}{0.96,0,0,0}
\usepackage{lipsum,adjustbox}

\usepackage[subnum]{cases}
\usepackage{color,soul}

\usepackage{subcaption}

\usepackage{footnote}
\makesavenoteenv{tabular} 

\usepackage{multirow}

\usepackage{booktabs,subcaption,amsfonts,dcolumn}
\newcolumntype{d}[1]{D..{#1}}

\usepackage{xcolor}

\usepackage{textcomp}

\title{Monolith to Microservices: Representing Application Software through Heterogeneous Graph Neural Network}

\author{
Alex Mathai$^1$
\and
Sambaran Bandyopadhyay$^2$\footnote{The work was done when Sambaran was affiliated to IBM Research prior to joining Amazon}\and
Utkarsh Desai$^{1}$\footnote{Utkarsh is currently at Google}\And
Srikanth Tamilselvam$^1$
\affiliations
$^1$IBM Research\\
$^2$Amazon
\emails
\{alexmathai98, samb.bandyo, utk.is.here, srikanthtamilselvam\}@gmail.com
}

\begin{document}

\maketitle

\begin{abstract}

Monolithic software encapsulates all functional capabilities into a single deployable unit. But managing it becomes harder as the demand for new functionalities grow. Microservice architecture is seen as an alternate as it advocates building an application through a set of loosely coupled small services wherein each service owns a single functional responsibility. But the challenges associated with the separation of functional modules, slows down the migration of a monolithic code into microservices. 
In this work, we propose a representation learning based solution to tackle this problem. We use a heterogeneous graph to jointly represent software artifacts (like programs and resources) and the different relationships they share (function calls, inheritance, etc.), and perform a constraint-based clustering through a novel heterogeneous graph neural network. Experimental studies show that our approach is effective on monoliths of different types. 

\end{abstract}

\section{Introduction}\label{sec:intro}
Monolith architecture is the traditional unified model for designing software applications. It encapsulates multiple business functions into a single deployable unit. But such applications become difficult to understand and hard to maintain as they age, as developers find it difficult to predict the change impact \cite{kuryazov2020towards}. Therefore, microservice \cite{fowler,thones2015microservices} architectures are seen as an alternative. It aims to represent the application as a set of small services where each service is responsible for a single functionality. It brings multiple benefits like efficient team structuring, independence in development \& deployment, enables flexible scaling and less restriction on technology or programming language preference. But migrating from monolith to microservices is a labour intensive task.  It often involves domain experts, microservices architects and monolith developers working in tandem to analyze the application from multiple views and identify the components of monolith applications that can be turned into a cohesive, granular service. They also need to work with constraints like the exact number of microservices to be exposed and components that should definitely be part of a particular microservice. 


The software engineering community refers to this migration process as a software decomposition task. Many works \cite{tzerpos2000accd,harman2002new,mazlami-icws-2017,mahouachi2018search,mancoridis1999bunch} 
leverage the syntactical relationships between the programs and treated this decomposition as an optimization problem to improve different quality metrics like cohesion, coupling, number of modules, amount of changes etc. While the accuracy of these approaches has been evolving over time, they have their drawbacks such as 1) reliance on external artifacts like logs, commit history etc. 2) focus on only a subset of the programs 3) less attention to non program artifacts like the tables, files 4) minimal consideration for transactional data. Recently \cite{jin2019service,kalia2020mono2micro} executed test cases to extract runtime traces. Each execution is considered as a business function and they try to cluster business functions. But this work relies on access to runtime traces and complete coverage of test cases which cannot be always guaranteed. Also, these work did not consider data entities for decomposition. 

\begin{figure}[!t]
  \centering
  \includegraphics[width=\linewidth]{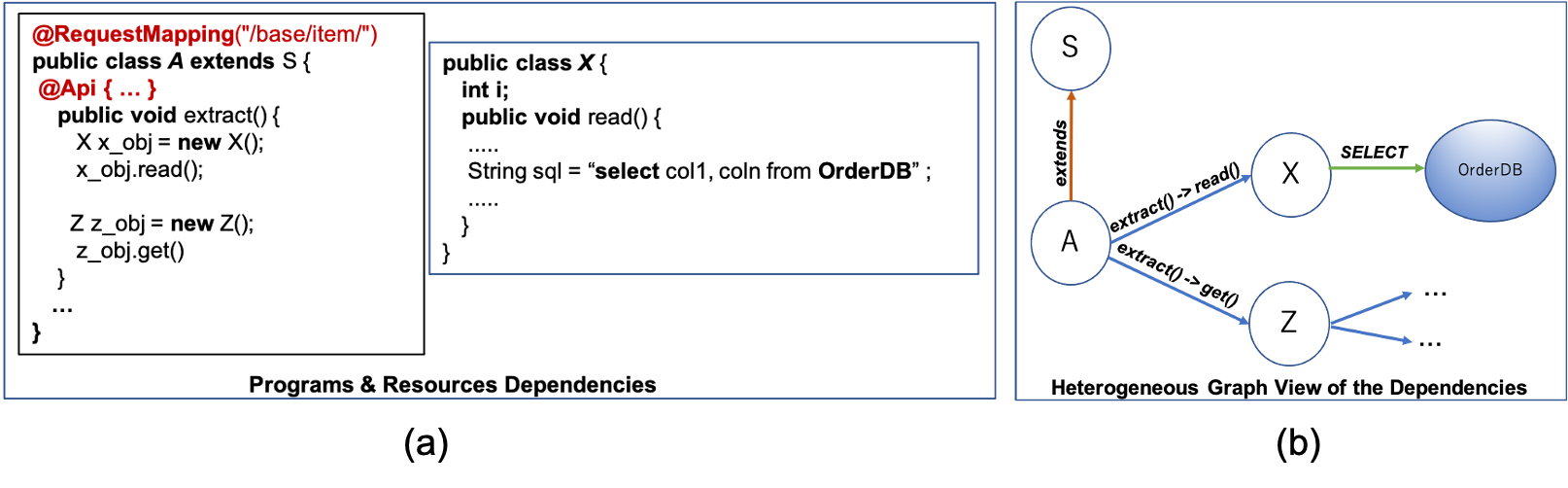}
 \caption{Representing software as a heterogeneous graph: (a) Shows a class \textit{A} that \texttt{extends} class \textit{S}. \textit{A} implements the \texttt{extract()} method where it invokes class \textit{X} \texttt{read()} method. The \texttt{extract()} method also has the \texttt{@Api} annotation indicating that it exposes a service. We also notice class \textit{X} performing a \texttt{READ} operation on table \textit{OrderDB}. (b) Shows the graph view with classes \textit{A}, \textit{S}, \textit{X} as \textit{program} nodes and table \textit{OrderDB} as a \textit{resource} node. Additionally, the different kind of dependencies are represented as edges $e(A,S)$, $e(A,X)$, $e(A,Z)$ and $e(X,OrderDB)$}.
  \label{fig:code_graph}
\end{figure}

Graphs are a natural choice to represent the application's structural and behavioral information \cite{mancoridis1998using,desai2021graph}. The structural information consisting of different application entities such as programs, files, database tables can be represented as nodes and their different relationships such as \textit{calls, extends, implements} between program to program and different CRUD operations that happen from program to data resources (table, file) can be represented as edges in the graph. Figure \ref{fig:code_graph} captures the construction of a heterogeneous graph from a sample java code. The behavioral information of the application identified through the sequence of programs and data resources that come together to support a business function can be captured as node/edge attributes. The monolith to microservices task can thus be viewed as a graph based clustering task which involves 1) Representation learning of application implementation from the graph structure and 2) Using this learnt representation for clustering. Graph neural networks have achieved state of the art results for multiple graph-based downstream tasks such as node classification and graph classification \cite{kipf2017semi,xu2018how,velivckovic2017graph}.
Most graph neural networks follow message passing mechanisms where the vector representation of a node is updated by combining its own features and aggregated features from its neighborhood. 
Recently \cite{desai2021graph} showed how the programs and its relationships in the application can be represented as a graph and proposed a multi-objective graph convolution network that combined node representation \& node clustering by diluting outliers. But since the framework did not consider application's data resources like database tables, files and the different relationships that exists between programs \& resources in the graph construction, the functional independence property of microservices is not completely satisfied. In addition, application architects have a \textit{functional view} of the application, so they decide on the target number of microservices and identify the core representative programs or tables from the monolith for each microservice. The solution should therefore accept the architect inputs as constraints and form clusters to maximize functional alignment.       
In this work, we propose a novel graph neural network based solution to refactor monolith applications into a desired number of microservices. The main contributions of our paper are listed below.
\begin{enumerate}
    \item We translate the application software's structural and behavioral properties into a  heterogeneous graph through nodes, edges and node/edge attributes.
    \item We introduce a novel heterogeneous graph neural network (GNN), referred to as CHGNN, that enables the representation of both \textit{data resources} and \textit{programs}. For the first time in literature, we perform a constraints-based clustering jointly in the framework of heterogeneous GNN. 
    \item We show that inclusion of heterogeneous information generates better quality microservices recommendations through four publicly available monolith applications.
\end{enumerate}

\section{Methodology}\label{sec:soln}


Given a monolith application and constraints $S_{list}$ -  the list of $K$ seed sets from the subject matter experts (SMEs), we want to partition the monolith into $K$ fairly distributed clusters where each cluster is a group of programs and resources that perform a well-defined functionality. 


\subsection{Converting Applications to Graph}\label{sec:mapToGraph}
We now describe our approach to represent an application as a graph. The primary programming construct in different languages is different - a class in Java and a program in COBOL. Hence, in the rest of this work, we refer to classes or programs as simply programs for consistency. Consider a simple Java application as shown in Figure \ref{fig:code_graph}. Each class or program in the application can be represented as a \textit{Program} node in the graph. Certain programs might also access resources such as database tables, files or other data structures. These can be represented as \textit{Resource} nodes in the graph. We denote the combined set of \textit{Program} and \textit{Resource} nodes as $V$, the set of all nodes. 
We establish an undirected \texttt{CALLS} edge from \textit{Program} node A to \textit{Program} node B if there is method in the program A that calls a method from program B. We also identify resource usage and create a \texttt{CRUD} edge from \textit{Program} node X to \textit{Resource} node R, if the program X accesses resource R. Static analysis tools\footnote{https://github.com/soot-oss/soot} can analyze the application code and identify the call chains and resource usage. $E$ denotes the combined set of all edges between the various nodes in the graph. Multiple method calls or resource usages between two nodes are still represented by a single unweighted edge.


We now generate the node attribute matrix, corresponding to the \textit{Program} and \textit{Resource} nodes of the graph. APIs exposed by applications are referred as EntryPoint Specifications \cite{dietrich2018driver}, or simply, \textit{Entrypoints} (EPs). The methods invoked through these APIs are annotated with tags such as \texttt{@Api} as shown in Figure \ref{fig:code_graph}. We refer to such methods as entrypoint methods and the corresponding programs as \textit{entrypoint programs}. Each entrypoint program can thus be associated with multiple entrypoints due to different entrypoint methods. 
From an entrypoint method, we can obtain a sequence of invoked methods and their corresponding programs using the execution traces of that Entrypoint. If $EP$ is the set of Entrypoints in an application and $V_P$, is the set of \textit{Program} nodes, we can define a matrix $A^{|V_P| \times |EP|}$, such that $A(i,p) = 1$ if program $i$ is present in the execution trace of entrypoint $p$, else $0$. Additionally, we define another matrix $C^{|V_P| \times |V_P|}$ such that $C(i,j)$ is the number of Entrypoint execution traces that contain both programs $i$ and $j$. If a program is not invoked in an execution trace for any Entrypoint, we remove the corresponding non-reachable \textit{Program} node from the graph. Finally, classes or programs may also inherit from other classes or programs or implement Interfaces. In Figure \ref{fig:code_graph}, class \textit{A} inherits from class \textit{S}. Although this establishes a dependency between the programs, it is not a direct method invocation. Hence, this dependency is not included as an edge in the graph, but as a \textit{Program} node attribute. Therefore, we define a matrix $I^{|V_P| \times |V_P|}$ and set $I(i,j) = I(j,i) = 1$ if programs $i$ and $j$ are related via an inheritance relationship and $0$ otherwise. The attribute matrix for \textit{Program} nodes is the concatenation of $A$, $C$ and $I$, and denoted as $X_P$.

For \textit{Resource} nodes, the Inheritance features are not applicable. The $A$ and $P$ matrices are obtained by summing up the corresponding rows from the respective $X_P$ matrices. The relationship between \textit{Program} nodes and \textit{Resource} nodes is many-to-many and this formulation simply aggregates features from all related programs into the resource to form the resource attribute matrix $X_R$. Each constituent matrix of $X_P$ and $X_R$ is row-normalized individually. The final set of node attributes are denoted as $X_V = \{X_P, X_R\}$.
The Edge attributes for \texttt{CALLS} edges is simply the vector $[1,0]$ and there are no additional features. For \texttt{CRUD} edges, the attribute vector represents the type of resource access performed. Since a program can access a resource in more than one fashion, this is a $1 \times 4$ vector, where each attribute represents the associated access type - [Create, Read, Update, Delete]. Hence a program that reads and updates a \textit{Resource} node will have $[0,1,1,0]$ as the edge feature. The edge attribute matrix is represented as $X_E$. 

Thus, an application can be represented by a heterogeneous graph as $G = (V, E, X_V, X_E)$. Let us assume that $\phi(v)$ and $\psi(e)$ denote the node-type of $v$ and edge-type of $e$ respectively. Let us use $\mathbf{x}_v \in \mathbb{R}^{D_{\phi(v)}}$ to denote the attribute vector for the node $v$ which belongs to $D_{\phi(v)}$ dimensional space. Similarly, $\mathbf{x}_e \in \mathbb{R}^{D_{\psi(e)}}$ is the edge attribute of the edge $e$.


\subsection{Proposed Heterogeneous Graph Neural Network}\label{sec:HGNN}
In this subsection, we aim to propose a graph neural network (GNN) which can (i) handle different node and edge types in the graph, (ii) obtain vector representation of both nodes and edges by jointly capturing both the structure and attribute information, (iii) output community membership of all the nodes in the graph in a unified framework. We refer the proposed architecture as \textit{CHGNN} (\underline{C}ommunity aware \underline{H}eterogeneous \underline{G}raph \underline{N}eural \underline{N}etwork). There are different steps in the design of CHGNN as described below.

\subsubsection{Mapping Entities to a Common Vector Space}\label{sec:mapToCommon}
Due to heterogeneity from different software artifacts, attributes associated with nodes and edges of the input graph are not of same types and they can have different dimensions. Such heterogeneity can be addressed in the framework of message passing heterogeneous graph neural networks in two ways: (i) Map the initial attributes to a common vector space using trainable parameter matrices at the beginning \cite{wang2019heterogeneous}; (ii) Use different dimensional parameter matrices while aggregating and combine information at each step of message passing \cite{Vashishth2020Composition-based}. We choose the first strategy since that makes the subsequent design of the GNN simpler and helps to add more layers in the GNN. So, we introduce type specific trainable matrices $W_{\phi(v)} \in \mathbb{R}^{F^{(0)} \times D_{\phi(v)}}$ for nodes and $W_{\psi(e)} \in \mathbb{R}^{F^{(0)} \times D_{\psi(e)}}$ for edges $\forall v \in V$ and $\forall e \in E$.
\begin{align}\label{eq:commonSpace}
    \mathbf{h}_v^{(0)} = \sigma(W_{\phi(v)} \mathbf{x}_v) ; \; \mathbf{h}_e^{(0)} = \sigma(W_{\psi(e)} \mathbf{x}_e)
\end{align}
where $\sigma$ is a nonlinear activation function. $\mathbf{x}_v$ and $\mathbf{x}_e$ are the initial attribute vectors of node $v$ and edge $e$ respectively. $\mathbf{h}_v^{(0)}$ and $\mathbf{h}_e^{(0)}$ are considered as $0$th layer embeddings for $v$ and $e$ respectively. They are fed to the message passing framework as discussed below.

\subsubsection{Message Passing Layers for Nodes and Edges}\label{sec:layer}
Message passing graph neural networks have achieved significant success for multiple node and graph level downstream tasks. In this framework, we obtain vector representation for both nodes and edges of the heterogeneous graph. There are $L \geq 1$ message passing layers. We define the $l$th layer ($1 \leq l \leq L$) of this network as follows. 

As the first step of a message passing layer, features from the  neighborhood are aggregated for each node. In recent literature, it has been shown that obtaining both node and edge representation improves the downstream performance for multiple applications 
\cite{jiang2019censnet,bandyopadhyay2019beyond}. Following that, we also design the GNN to exchange features between nodes and edges, and update the vector representation for both. In each layer, we use two parameter matrices $W_1^{(l)} \in \mathbb{R}^{F^{(l)} \times F^{(l-1)}}$ and $W_2^{(l)} \in \mathbb{R}^{F^{(l)} \times F^{(l-1)}}$ to handle node and edge embeddings respectively. For a node $v \in V$, its neighborhood information is aggregated as:
\begin{equation}\label{eq:aggre}
    \mathbf{z}_v^{(l)} = \sum\limits_{u \in \mathcal{N}(v)} \frac{1}{\sqrt{d_u} \sqrt{d_v}} W_1^{(l)} \mathbf{h}_u^{(l-1)} * \sigma\big( W_2^{(l)} \mathbf{h}_{uv}^{(l-1)} \big)
\end{equation} 
where $d_u$ and $d_v$ are the degrees of the nodes $u$ and $v$ respectively. $\frac{1}{\sqrt{d_u} \sqrt{d_v}}$ is used to symmetrically normalize the degrees of the nodes in the graph \cite{kipf2017semi}. $\sigma()$ is a nonlinear activation function and $*$ is Hadamard (element-wise) product. Next, the aggregated information $\mathbf{z}_v^{(l)}$ is combined with the embedding of node $v$ to update it as follows.
\begin{equation} 
\label{eq:nodeEmb}
    \mathbf{h}_v^{(l)} = \sigma \big(\mathbf{z}_v^{(l)}  + \frac{1}{d_{v}}\mathbf{h}_v^{(l-1)} \big)
\end{equation} 
$\mathbf{h}_v^{(l)}$ is considered as the node embedding of node $v$ at $l$th layer. To update the embedding of an edge $(u,v)$, we use the updated embeddings of two end point nodes and the existing embedding of the edge as follows ($||$: concatenation of vectors):
\begin{equation}
\label{eq:edgeEmb}
    \mathbf{h}_{uv}^{(l)} = \sigma \Big( W_3^{(l)} \big( \frac{\mathbf{h}_u^{(l)} + \mathbf{h}_v^{(l)}}{2} \; || \; \mathbf{h}_{uv}^{(l-1)} \big) \Big)
\end{equation} 
where $W_3^{(l)} \in \mathbb{R}^{F^{(l)} \times (F^{(l)} + F^{(l-1)})}$ is a parameter matrix. This completes the definition of the layer $l$ of the message passing network. Please note that the dimensions of the trainable matrices $W_1^{(l)}$, $W_2^{(l)}$ and $W_3^{(l)}$ determine the dimension of the embedding space of the nodes and edges.

To build the complete network, we first map the heterogeneous nodes and edges to a common space using Equation \ref{eq:commonSpace}. Subsequently, we use 2 message passing layers ($l = 1,2$) as encoders (compressing the feature space) and next 2 message passing layers ($l = 3,4$; $L=4$) as decoders (decompressing the feature space), with $F^{(0)} > F^{(1)} > F^{(2)} < F^{(3)} = F^{(1)} < F^{(4)} = F^{(0)}$. To map the node and edge features to their respective input attribute space $\mathbb{R}^{D_\phi(v)}$ and $\mathbb{R}^{D_\psi(e)}$, we again use linear transformations followed by activation functions as shown below.
\begin{align}\label{eq:invCommonSpace}
    \hat{\mathbf{x}}_v = \sigma( \hat{W}_{\phi(v)} \mathbf{h}_v^{(L)}) 
    \; ; \;\;
    \hat{\mathbf{x}}_e = \sigma( \hat{W}_{\psi(e)} \mathbf{h}_e^{(L)}) 
\end{align}
These reconstructed node and edge attributes are used to design the loss functions as discussed next.


\subsubsection{Design of the Loss Functions and Joint Clustering of Heterogeneous Nodes}\label{sec:clus}
We use three types of unsupervised reconstruction losses.

\textbf{Node Attribute Reconstruction}: We try to bring the initial node features $\mathbf{x}_v$ and the reconstructed node features $\hat{\mathbf{x}}_v$ close to each other by minimizing $ \sum\limits_{v \in V} || \mathbf{x}_v - \hat{\mathbf{x}}_v ||_2^2 $.

\textbf{Edge Attribute Reconstruction}: With similar motivation as above, we minimize $ \sum\limits_{e \in E} || \mathbf{x}_e - \hat{\mathbf{x}}_e ||_2^2$.

\textbf{Link Reconstruction}: Above two loss components do not capture anything about the link structure of the heterogeneous graph. Let us introduce the binary variables $a_{u,v}$ $\forall u, v \in V$ such that $a_{u,v}=1$ if $(u,v) \in E$ and $a_{u,v}=0$ otherwise. We want to ensure that embeddings of two nodes are close to each other if there is an edge between them by minimizing $ \sum\limits_{u,v \in V} \big( a_{u,v} - \mathbf{x}_u \cdot \mathbf{x}_v \big)_2^2 $.

\textbf{Unifying Node Clustering}: After we map the monolith to a heterogeneous graph, we cluster the nodes of the graph to form microservices.
As the nodes are represented in the form of vectors through the heterogeneous GNN encoder as discussed in Section \ref{sec:layer}, 
we unify the clustering objective with the heterogeneous GNN as follows.

The node embeddings at the end of encoding layers (i.e., $L/2$ layers) are $\mathbf{h}_v^{(L/2)}$, $\forall v \in V$. We design a k-means++ objective \cite{arthur2006k} by introducing two parameter matrices $M \in \mathbb\{0,1\}^{|V| \times K}$ and $C \in \mathbb{R}^{K \times F^{(L/2)}}$. $M$ is the binary cluster assignment matrix where each row sums up to $1$. We assume to know the number of clusters $K$. $M_{vk} = 1$ if node $v$ belongs to $k$th cluster and $M_{vk} = 0$ otherwise\footnote{To avoid cluttering of notations, we use $M_{vk}$ instead of $M_{\text{index}(v) k}$, where $1 \leq \text{index}(v) \leq |V|$}. $k$th row of $C$, denoted as $C_k$, is the center of $k$th cluster in the embedding space. Node clusters and the corresponding cluster centers can be obtained by minimizing clustering loss ($CL$) which is $\sum\limits_{v \in V} \sum\limits_{k=1}^K M_{vk} || \mathbf{h}_v^{(L/2)} - C_k ||_2^2$.

\textbf{Allowing seed constraints} : 
As motivated in Section \ref{sec:intro}, real-world applications generally have constraints provided by SMEs in the form of clustering seeds.
To incorporate such constraints, we take a list of seed sets $S_{list} = [S_1,\cdots , S_K]$ as input, where each $S_i$ is a set of seed nodes that must belong to the corresponding cluster $\mathcal{C}_i$. 
\begin{equation}
    \mathbf{SC}_{k} = \sum\limits_{i \in S_{k}} \mathbf{h}_{i}^{(L/2)} / |S_{k}|, \forall k \in [1,\cdots,K]
\end{equation}
\begin{equation} \label{eq:dist}
    Dist = -\sum_{k=1}^{K} \sum_{k'=1}^{K} || \mathbf{SC}_{k} -  \mathbf{SC}_{k'} ||_{2}^{2}
\end{equation}
With $S_{list}$ as input, we need to ensure two requirements - 1) In the final output, every seed set ($S_{i}$) must belong to a pre-determined cluster ($\mathcal{C}_i$) and 2) the embeddings of the seed sets in different clusters must be as far apart as possible. To address the first requirement we add hard constraints to our cluster assignment algorithm . This can be seen in Algorithm \ref{alg:CHGNN} (Lines $15$-$17$), where for each seed artifact, we assign the cluster manually. To address the second requirement we add soft constraints to our clustering loss function. This is captured by measuring the distance between the seed set centers ($\mathbf{SC}_{k}$) as shown in Equation \ref{eq:dist}. As the distance should be maximised, we negate this distance to maintain a minimization objective. 

Hence, the total loss to be minimized by CHGNN is: 
\begin{equation}
\label{eq:totalLoss}
\begin{aligned}
	\underset{\mathcal{W},M,C}{\text{min}} \; 
	\mathcal{L} = \alpha_1 \; \sum\limits_{v \in V} || \mathbf{x}_v - \hat{\mathbf{x}}_v ||_2^2 \;
	+ \alpha_2 \; \sum\limits_{e \in E} || \mathbf{x}_e - \hat{\mathbf{x}}_e ||_2^2 \\
	+ \alpha_3 \; \sum\limits_{u,v \in V} \big( a_{u,v} - \mathbf{x}_u \cdot \mathbf{x}_v \big)_2^2 + \alpha_4 \; (CL + Dist)
\end{aligned}
\end{equation}
where $\mathcal{W}$ contains the trainable parameters of the GNN described in Section \ref{sec:mapToCommon}. $\alpha_1$, $\alpha_2$, $\alpha_3$ and $\alpha_4$ are non-negative weights. We set them such that individual loss components contribute equally in the first iteration of the algorithm.

\subsection{Training and Analysis}
\begin{algorithm}[t] 
  \small
  \caption{\textbf{CHGNN}}
  \label{alg:CHGNN}
\begin{algorithmic}[1]
	\Statex \textbf{Input}: Class dependencies and $S_{list}$ (list of $K$ seed sets)
	\State Convert the application to a graph representation as defined in Section \ref{sec:mapToGraph} and obtain the $V$, $E$ and $X_{V}$ and $X_{E}$.
    \State Let $S_{list}$ be $[S_1, \cdots, S_K]$. Hence each cluster ($\mathcal{C}_{k}$) has a mandatory set of seeds ($S_{k}$).
    \State Let the list of clusters be $[\mathcal{C}_1, \cdots ,\mathcal{C}_K]$ and the list of centers for each cluster be $[C_{1},\cdots ,C_{K}]$.
    \State Pre-train the heterogeneous GNN encoder and decoder (refer \textbf{Training Procedure} in supplementary material)
    \State \% Initialize the cluster centers as follows \%
    \State $C_{k}$ = $\sum\limits_{k \in S_{k}} \mathbf{h}_{k}^{L/2} / |S_{k}|$, $\forall k \in [1,\cdots,K]$
	\For{T iterations}
	    \State \% Assign a cluster to each node \%
    	\For{$v \in V$}
    	    \If{$v$ not in $S_{list}$}
    	        \State \% Find closest centres for non-seeds \%
    	        \State Find the closest center ($C_{k}$) with Equation \ref{eq:clusAssign}  
    	        \State Add $v$ to the corresponding cluster $(\mathcal{C}_{k})$
    	   \Else
    	        \State \% Fix the cluster for seeds \%
    	        \State Find $S_{k}$ such that $v \in S_{k}$ 
    	        \State Add $v$ to the corresponding cluster $(\mathcal{C}_{k})$
    	    \EndIf
    	\EndFor
        \State Update cluster centers with Equation \ref{eq:clusCen}.
        \State Update the parameters of the heterogeneous GNN encoder and decoder by minimizing Eq. \ref{eq:totalLoss} using ADAM.
    \EndFor
    \Statex \textbf{Output}: Clusters and Cluster Centers
	\end{algorithmic}
  \end{algorithm}
First, we pre-train the parameters of the GNN without including the clustering loss component, i.e., setting $\alpha_4 = 0$ in Equation \ref{eq:totalLoss}. We use ADAM optimization technique 
to update the parameters of GNN. Once the pre-training is completed, we use alternating optimization techniques to update each of clustering parameters $M$ and $C$, and parameters of GNN $\mathcal{W}$, while keeping others fixed. Using Lloyd's update rule for k-means, 
we update $M$ and $C$ as:
\begin{equation} \label{eq:clusAssign}
    M(v,k)=
\begin{cases}
1, \; \text{if\;} k = \underset{k' \in \{1,\cdots,K\}}{\text{argmin}} ||\mathbf{h}_v^{(L/2)} - C_{k'} ||_2^2\\
0, \; \text{Otherwise}
\end{cases}
\end{equation}
\begin{equation} \label{eq:clusCen}
    C_{k} = \frac{1}{N_k} \sum\limits_{v \in \mathcal{C}_k} \mathbf{h}_v^{(L/2)}
\end{equation}


where $N_k = \sum\limits_{v \in V} M_{vk}$. Due to the presence of clustering loss component in Equation \ref{eq:totalLoss}, updating the parameters $\mathcal{W}$ of GNN can pull the node embeddings close to their respective cluster centers further, along with reconstructing initial node and edge attributes and the link structure.

      


\section{Experimental Evaluation}
\label{sec:eval}

 To study the efficacy of our approach, we chose four publicly-available monolith applications (links in supplementary material) namely Daytrader, PlantsbyWebsphere (PBW), Acme-Air and GenApp. Together, these applications show a good diversity in terms of the programming paradigms, languages and technologies used. Details of each monolith are provided in Table \ref{tab:metrics}. We did not include DietApp (used in baseline \cite{desai2021graph}), as the public repository does not expose the code used for interfacing between programs and tables. Hence DietApp reduces to a homogeneous graph (only programs). As expected, upon experimentation, we observe that CHGNN gets the same output as the baseline.
 
\subsection{Constraints to clustering}
For each application, we get the target number of microservices (\textit{K clusters}) and constraints of what entities each microservice (\textit{K seed sets}) should contain as inputs from the SMEs. Naturally, the seed sets cannot be shared/overlap with each other. Typically, the constraints include tables and program entities that are central to the cluster. They act as a means to guarantee functional alignment.

\subsection{Quantitative Metrics}
\label{sec:metrics}

To quantitatively evaluate the clusters, we use four established graph metrics. We briefly touch upon these below.

\textbf{1. Modularity (Mod)} : The Modularity metric \cite{modularity1} is widely used to evaluate the quality of generated graph partitions. It computes the difference between the \textit{actual} intra-edges of a cluster and the \textit{expected} intra-edges of the cluster in a randomly re-wired graph. Higher the modularity - the better the partitions.

\textbf{2. Non-Extreme Distribution (NED)} : The NED metric \cite{NED1} examines if the graph partitions are tiny, large or of an acceptable size. Rather than fixing the low and high limits to $5$ and $20$ \cite{NED1}, we take the average cluster size as input and check if the cluster size lies within the $50\%$ tolerance bandwidth of the average. This change helps NED generalise well to small and large sized applications.
\begin{equation} 
\label{eq:ned}
    NED =  \frac{1}{N} \sum\limits_{k=1}^{K} n_{k}, \forall n_{k} \in [(1-\epsilon)*\frac{N}{K},(1+\epsilon)*\frac{N}{K}]
\end{equation}

where $N$ and $K$ are the number of graph nodes and the number of clusters respectively. Here, we set $\epsilon$ to $0.5$.

\textbf{3. Coverage} :  Coverage \cite{2010} tries to measure \textit{cohesion} by computing the ratio of the number of intra-cluster edges to the total number of edges in the graph. Higher coverage implies higher cohesion which results in better clusters. 

\textbf{4. S-Mod} : S-Mod \cite{jin2019service} is another quantitative approach that measures the quality of generated partitions. It is computed by subtracting \textit{coupling} from \textit{cohesion}. More cohesion implies more intra-cluster edges and more coupling implies more inter-cluster edges. For microservices, it is ideal to have high cohesion and low coupling. Hence, higher S-Mod values are favourable.

\subsection{Combined Metric}\label{sec:tradeoff}
However in our experiments, we observe that a few \textit{boulder} (very large) clusters and many \textit{dust} (very small) clusters can increase intra-cluster edges and decrease inter-cluster edges. This improves S-Mod and Coverage considerably, but leads to a poor NED score - which penalises for size imbalance. Hence, instead of relying on a single metric, we calculate $Metric_{Sum}$ (summing all metrics) and rank our approaches accordingly. $Metric_{Sum} \in [-2,4]$ as Mod $\in [-1,1]$, NED $\in [0,1]$, S-Mod $\in [-1,1]$ and Coverage $\in [0,1]$. In all our experiments, we observe that every metric is positive for all the algorithms considered.

\begin{figure*}[t]
     \centering
     \resizebox{0.83\linewidth}{!} {
          \begin{subfigure}[b]{0.5\textwidth}
             \centering
             \includegraphics[width=\textwidth]{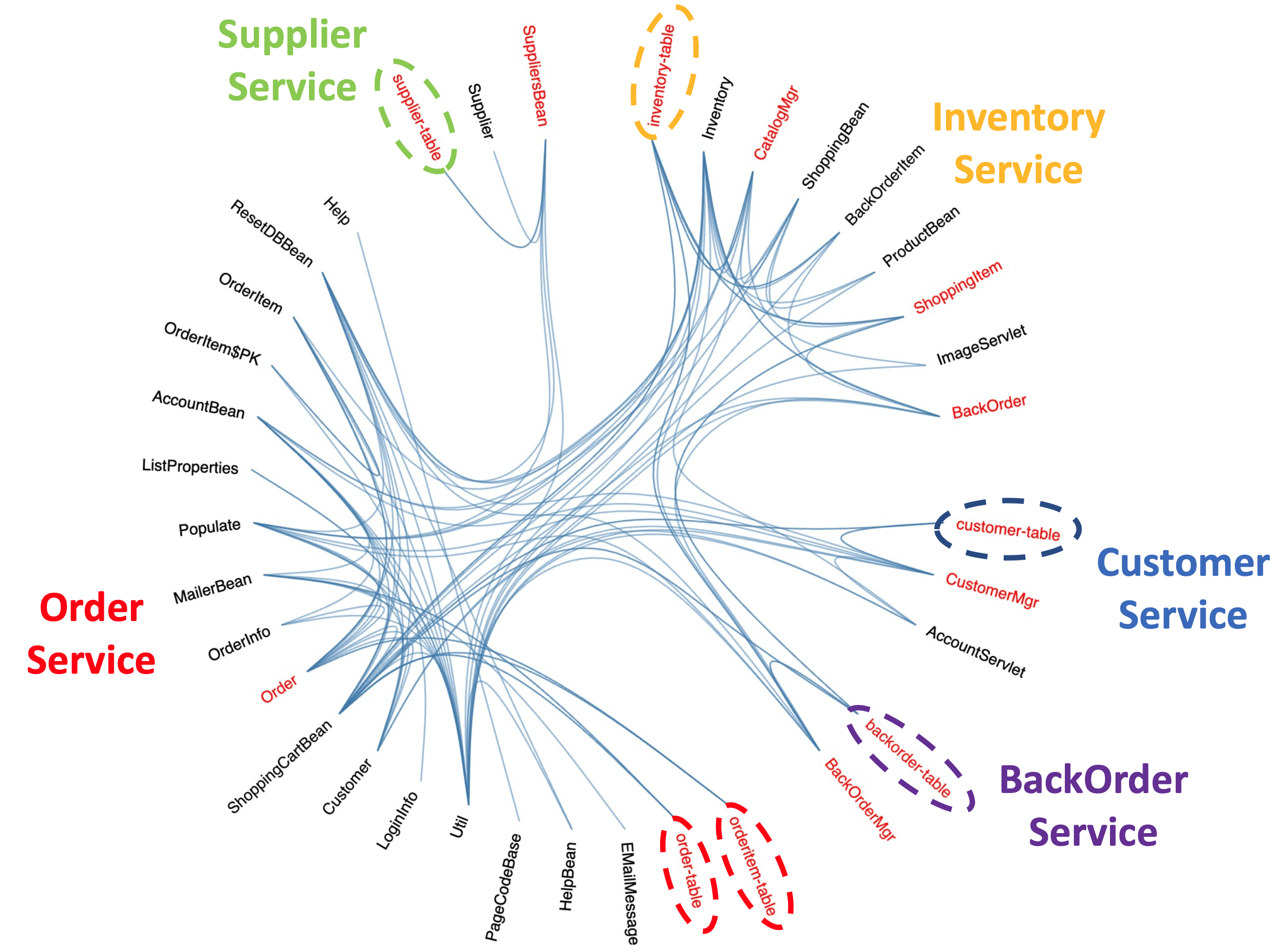}
             \caption{Results for COGCN++}
             \label{fig:COGCN}
         \end{subfigure}
         \hfill
         \hfill
         \begin{subfigure}[b]{0.45\textwidth}
             \centering
             \includegraphics[width=\textwidth]{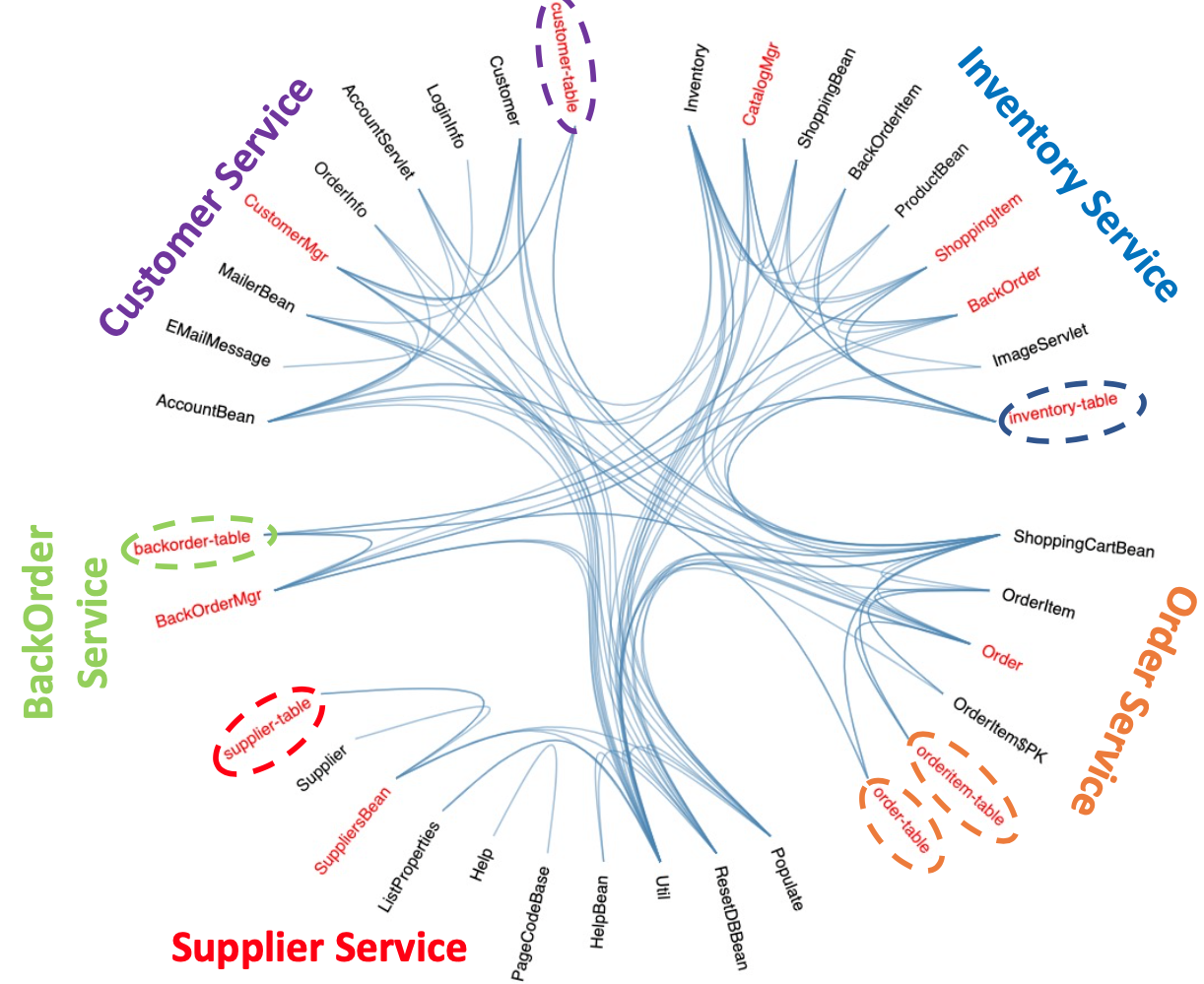}
             \caption{Results for CHGNN}
             \label{fig:CHGNN}
         \end{subfigure}
     }
        \caption{Candidate microservices for PBW application : The clusters capture the five business functions of PBW. The bounding circles indicate the different \textit{resources} (\textit{tables}) owned by the clusters.The nodes in \textit{red} fonts are seeds provided by SME.}
        \label{fig:pbwclusters}
\end{figure*}

\subsection{Baseline Algorithms and Experimental Setup} \label{sec:setup}
As converting a monolith application to a heterogeneous graph neural network and applying them for clustering is a novel direction, we have designed most of the baselines with motivations from existing works on graph representation learning. They are discussed below.

\textbf{COGCN++} : \cite{desai2021graph} introduced COGCN, wherein, the monolith application (having programs and resources) is converted to a homogeneous graph. Subsequently, COGCN, which has GCN layers trained on reconstruction and clustering loss, is applied on the homogeneous graph to obtain the micro-services. When running vanilla COGCN, we noticed that it does not ensure the mutual exclusivity of seed sets. Hence we add the seed constraint loss (Equation \ref{eq:dist}) to fulfill this requirement. We refer to this model as COGCN++. Note : COGCN++ does not consider edge embeddings/attributes - it only considers node embeddings.

\textbf{HetGCNConv}: Here, we create the heterogeneous graph as discussed in Section \ref{sec:mapToGraph}, but consider all the edges as of similar types. We map the heterogeneous nodes to a common vector space as done in Section \ref{sec:mapToCommon} by using node-type specific parameter matrices $W_{\phi(v)}$'s. We then use the GCN convolution model and train it by setting $\alpha_2$ to $0$. 

\textbf{CHGNN-EL} : A variant of our proposed CHGNN where we drop the edge feature re-construction loss from the optimization procedure by setting $\alpha_2=0$ in Equation \ref{eq:totalLoss}.

\textbf{CHGNN} : This is our final model proposed in this work.

Please note that COGCN++ is a homogeneous graph based approach. HetGCNConv and CHGNN-EL are two variants of our proposed heterogeneous model CHGNN. HetGCNConv only uses heterogeneous nodes but homogeneous edges. CHGNN-EL uses both heterogeneous nodes and edges, but sets $\alpha_2$ to $0$. For all relevant architecture specifications, training procedures and hardware requirements, refer to the supplementary material.

\subsection{Qualitative and Quantitative Results}\label{sec:results}

For our qualitative analysis, we study our predictions for the PBW application and compare our results with COGCN++. As seen in Figure. \ref{fig:CHGNN}, we observe that CHGNN using the seed inputs ($S_{list}$) has identified five functional clusters - Customer, Order, Inventory, BackOrder and Supplier. The programs in each cluster have very close dependencies within the cluster and contribute majorly to a common business function. This is also evident from the similar names of artifacts in a cluster (like \textit{OrderItem}, \textit{Order-table}, \textit{Orderitem-table} in the \textit{Order} service). Associated with each cluster is a group of data nodes (\textit{dashed circled}) that interact closely with the programs in their cluster. As seen, the \textit{Supplier} cluster has \textit{supplier-table}, \textit{Supplier}, \textit{SupplierBean} and \textit{Populate} nodes. In Figure \ref{fig:COGCN}, we observe that COGCN++ is successful at separating seed sets, as it leverages the seed constraint loss. However, we notice the following differences. 

1. Unlike CHGNN that creates evenly sized clusters, COGCN++ creates a few \textit{boulder} clusters like the \textit{Order} cluster (having $19$ artifacts) and many \textit{dust} clusters like \textit{Supplier}, \textit{BackOrder} and \textit{Customer} (having atmost $3$ artifacts). As explained in Section \ref{sec:tradeoff}, this strategy helps COGCN++ outperform on S-Mod and Coverage but dramatically underperform on NED. For the task of microservices partitioning, having skewed cluster sizes may result into few over-utilized services and many under-utilized services - which is not desired.

2. We also find that COGCN++ wrongly associates many artifacts that actually belong to the \textit{Customer} service (like \textit{AccountBean}, \textit{LoginInfo} and \textit{EmailMessage}) with the \textit{Order} and \textit{Inventory} service. This is not the case for CHGNN.

All of the above observations have been validated by one of our SMEs (R1) - {\say{CHGNN is better. Reason : The clusters are more evenly distributed. Customer service came out well with account management (AccountBean, LoginInfo, EmailMessage) contained in it which is desired.}}

\begin{table}[ht]
\resizebox{1.0\linewidth}{!} {
\begin{tabular}{c|c|c|ccc}
\toprule
Dataset     & \multicolumn{1}{c|}{Details}  & \multicolumn{1}{c|}{} & \multicolumn{3}{c}{HET-GNN Variations}                             \\
&  & COGCN++ & HetGCNConv     & CHGNN-EL       & CHGNN          \\ \hline
\begin{tabular}[c]{@{}c@{}}ACME\\ (Airline App,\\ Lang: Java)\end{tabular}       & \begin{tabular}[c]{@{}c@{}}K=4\\\#Class=30\\\#Resource=6\end{tabular}   & 1.329   & 1.712                                                 & 1.775          & \textbf{1.784} \\ \hline
\begin{tabular}[c]{@{}c@{}}DayTrader \\ (Trading App, \\ Lang:Java)\end{tabular} & \begin{tabular}[c]{@{}c@{}}K=6\\ \#Class=111\\ \#Resource=11\end{tabular} & 1.156   & 1.096                                                 & 1.310          & \textbf{1.336} \\ \hline
\begin{tabular}[c]{@{}c@{}}PBW\\ (Plant Store, \\ Lang:Java)\end{tabular}        & \begin{tabular}[c]{@{}c@{}}K=5\\ \#Class=30\\ \#Resource=6\end{tabular}   & 1.583   & 1.577                                                 & \textbf{1.805} & 1.762          \\ \hline
\begin{tabular}[c]{@{}c@{}}Genapp\\ (Insurance App, \\ Lang:Cobol)\end{tabular}  & \begin{tabular}[c]{@{}c@{}}K=4\\ \#Class=30\\ \#Resource=10\end{tabular}  & 1.870   & \textbf{2.016}                                        & 2.010          & 2.000\\ \bottomrule
\end{tabular}}
\caption{Performance of partitioning Monoliths (average of $30$ runs)}
\label{tab:metrics}
\end{table}

We depict our quantitative results in terms of $Metric_{Sum}$ in Table \ref{tab:metrics} and explain two trends that we have observed.

(i) COGCN++ scores lower values for $Metric_{Sum}$ consistently in every application. Hence, on an average, the heterogeneous graph formulations outperform COGCN++. 

(ii) In all the applications, either CHGNN or CHGNN-EL appear in the top two results for $Metric_{Sum}$.
Hence, on an average, both models are relatively consistent across the varying application topologies.


\textbf{Summary}: From this, it is evident that (1) Heterogeneous graph formulations always guarantee better performance as can be seen in the three HET-GNN variations in Table \ref{tab:metrics}.
(2) Clusters in CHGNN are evenly distributed and more meaningful in nature when compared to COGCN++. 

For extensive qualitative and metric-specific quantitative analysis of each application, please refer to the supplementary material.

\section{Conclusion}\label{sec:conclusion}
We proposed a novel heterogeneous GNN that enables representation of application data resources and programs jointly for recommending microservices. Both quantitative and qualitative studies show the effectiveness of heterogeneous graph formulations. In the future, we aim to study the decomposition task at a more granular level from \textit{programs} to \textit{functions} and \textit{tables} to \textit{columns}.

\bibliographystyle{named}
\bibliography{references}

\section{Supplementary Material}


\subsection{Hardware Requirements}\label{sec:hard_req}
All experiments were run on a system with 32 GB RAM, a 6-Core Intel i7 processor and a 4 GB AMD Graphics Card.


\subsection{Notations Used}
We summarize the notations in Table \ref{tab:notations}.

\begin{table}[H]
\centering
\resizebox{\linewidth}{!}{%
\begin{tabular}{*6c}
	\toprule
	\sffamily{Notations} & Explanations\\
    \hline
	\midrule
	$G = (V,E,X_V, X_E)$ & Input heterogeneous graph \\
	$e = (u,v)$ & An edge between two nodes \\
	$X_V$ & Set of node features \\
	$X_E$ & Set of edge features \\
	$\phi(v)$ and $\psi(e)$ & Node and edge types respectively\\
	$\mathbf{x}_v \in \mathbb{R}^{D_{\phi(v)}}$ & Initial attribute vector for the node $v$ \\
	$\mathbf{x}_e \in \mathbb{R}^{D_{\psi(e)}}$ & Initial attribute vector for the edge $e$ \\
	$\mathbf{h}_v^{(l)}$ & Embedding of node $v$ in layer $l$ \\
	$\mathbf{h}_e^{(l)}$ & Embedding of edge $e$ in layer $l$ \\
	$\hat{\mathbf{x}}_v$ & Reconstructed attribute vector for the node $v$ \\
	$\hat{\mathbf{x}}_e$ & Reconstructed attribute vector for the edge $e$ \\
    \bottomrule
	\end{tabular}
	}
\caption{Different notations used in the paper}
\label{tab:notations}
\end{table}

\subsection{Architecture Specifications}
In each of the experiments, we use a two-layer encoder and a two-layer decoder. The two encoder layers reduce the dimensions from  $F^{(0)}$ to $64$ and $32$ respectively. Similarly, the two decoders increase the dimensions from $32$ to $64$ and then from $64$ to $F^{(0)}$. 

\subsection{Training Procedure}
Before adding the clustering loss component, we pre-train the model for $150$ epochs ($lr=0.01$), allowing the model to better understand the application's graph structure. For this first round of training, we set $\{\alpha_1, \alpha_2, \alpha_3, \alpha_4 \}=\{0.4, 0.2, 0.4, 0\}$ for CHGNN. Similarly, we set $\{\alpha_1, \alpha_2, \alpha_3, \alpha_4 \} = \{0.5, 0, 0.5, 0\}$ for the others. We then add the clustering loss and train the model for another $150$ epochs ($lr=0.005$). During this second round of training, we set $\{\alpha_1, \alpha_2, \alpha_3, \alpha_4 \} = \{0.1, 0.1, 0.1, 0.7\}$ for CHGNN. Similarly, we set $\{\alpha_1, \alpha_2, \alpha_3, \alpha_4 \} = \{0.1, 0, 0.1, 0.8\}$ for the others. To generate the final cluster assignments for each application, we use the latest values of $M(i,k)$ - i.e. the value at the end of the $300^{th}$ epoch. 

\subsection{Time Complexity of CHGNN}
The forward pass of heterogeneous GNN takes $O(|E|)$ time since messages are computed and passed over the edges of the graph. Link reconstruction component in Section 2.2 of the main paper takes $O(|V|^2)$ time. This can easily be relaxed by reconstructing only the existing nodes (i.e., when $a_{u,v} = 1$) with some negative samples for non-existing edges 
.Since the number of nodes in the constructed heterogeneous graph which represents the monolith, is typically not very large for most real world applications, we reconstruct the full link structure.

\subsection{Links to Data and Toolkits Used} 
The following public Applications - Daytrader \footnote{https://github.com/WASdev/sample.daytrader7}, Plantsbywebsphere (PBW) \footnote{https://github.com/WASdev/sample.plantsbywebsphere}, Acme-Air\footnote{https://github.com/acmeair/acmeair} and GenApp \footnote{https://www.ibm.com/support/pages/cb12-general-insurance-application-genapp-ibm-cics-ts} are used for this study. All of them are Apache Licensed assets. We used the PyTorch Geometric \footnote{https://github.com/rusty1s/pytorch-geometric} framework for model implementation which is released as MIT License.

\subsection{Qualitative Study}\label{sec:study}
In this section, we cover the job profile of the participants, details on how the study was conducted and the feedback on the microservices recommendations. We also provide a detailed comparative analysis on three of the applications covering two programming paradigms.

\subsubsection{Participants Profile}\label{sec:profile}
To study the efficacy of the microservices recommendations, we requested participation from four software engineers to analyze four applications. On an average, the participants had industrial experience of 13 years in different software engineering roles. All the four participants had prior working experience on Java programming language. Two of the participants also had experience with working on COBOL applications. Two annotators (R1 \& R2) who had an understanding of COBOL took an average of 4 weeks to understand the GenApp application before they participated in this study. Two other annotators (R3 \& R4) spent an average of 2 weeks to understand the three Java based applications before they participated in this study.

\subsubsection{Study Instructions}
For each application, we provided the instructions shown in Figure. \ref{fig:instructions} to the respective participants. The application specific details like the application code reference and clustering outputs (the microservices recommendations - available as jsons and sunburst chart images) are mentioned in each instruction. As an example, Figure \ref{fig:acme-air comparison} captures the sunburst chart for the CHGNN and COGCN++ approaches that were presented to R3 \& R4.

\begin{figure}[ht]
\centering
\begin{minipage}[b]{0.9\linewidth}
\textit{
We thank you for agreeing to evaluate our work. Please go through the monolith application (https://www.ibm.com/support/pages/cb12-general-insurance-application-genapp-ibm-cics-ts) to make yourself familiar with the implementation structure. In addition, we are available to give an overview of the application capabilities and explain how the function works through different artifacts in the application. Below we have provided two sunburst charts, namely genappblind1.png and genappblind2.png showing the microservices recommendations for GenApp as clusters. Cluster boundaries can be identified by the big gap between them. Each label is a program/resource artifact in the GenApp application. To differentiate between the two, resource names have a suffix like “-table, -res or -db2table", or else, a prefix like "db2- or vsam-". The lines going across indicate the dependencies between the nodes. Incase you find text to be an easy way to understand the clusters, we have additionally, provided two json files genappblind1.json and genappblind2.json corresponding to the two sunburst charts. Please go over the two different results and tell us which clustering output seems to be more modular/independent and whether the artifacts within each cluster are closely related. Additionally, you can also mention the reasons behind your selection.}
\caption{Sample qualitative study instruction for the participant. Details like code repository link, cluster images and jsons are updated for the different applications}
\label{fig:instructions}
\end{minipage}
\end{figure}

\begin{table*}[t]
\small
\centering
\begin{tabular}{|p{0.1\textwidth}|p{0.1\textwidth}|p{0.1\textwidth}|p{0.6\textwidth}|}\hline
\textbf{Application} & \textbf{Language} & \textbf{Reviewer Selection} & \textbf{Reviewer Reasons} \\
\hline
\multirow{1}{*}{GenApp} & \multirow{1}{*}{COBOL} &  \multirow{1}{*}{CHGNN}  &  
 
 R2.  The main difference between Blind 2 and Blind 1 is which cluster the Business Rules belongs to. At the outset - considering that either Customer process or the Policy process will have to access these Business Rules wherever it gets added - it doesn't make much of a difference. But in a real-world engagement, it does make a difference because the sizing of the clusters have impact on the planning and implementation, especially since incremental and iterative implementation is preferred.
So, considering that the cluster sizing is more balanced in Blind 2 whereas in Blind 1, the cluster sizing is more skewed, it is preferable to go with Blind 2. More balanced cluster sizing means more equally distributed implementation cycles and possibly less rework on the co-existence architecture.

\\\cline{3-4}
\hline
\multirow{2}{*}{Daytrader} & \multirow{2}{*}{Java} &  CHGNN  & R3. I choose Cluster 2. Recommendations are evenly distributed compared to cluster1. In Cluster 2, holdingejb-table which is accessed only by TradeDirect/TradeSLSBean is grouped with one its dependency. I feel it is justified that TradeDirect is with accountejb-table and not with holdingejb- table. Cluster 2 also looks more functionally aligned,  Orderservices has come out with OrderData, orderejb-table, orders filter,orderdatabean. Similarly Account services has its most dependent artifacts in Cluster 2 compared to Cluster 1.
\\\cline{3-4}
& & CHGNN & R4.	Blind 1 :  TradeSLSBean and TradeDirect are clubbed in the same cluster along with AccountDataBean which looks very out of place. Also, the large cluster which contain more than 50\% of the classes could be better refactored. There are also some isolated classes and resources which could have been better arranged.

Blind 2 : This is not necessarily a very good clustering, but it still groups classes much better. There is a clear separation of functionality for OrderData and AccountData. The clusters are also somewhat evenly sized.
\\\cline{3-4}
\hline
\multirow{2}{*}{PBW} & \multirow{2}{*}{Java} &  CHGNN  & R3. Cluster 2  is  better.  Reason  :  The  clusters are more evenly distributed. Customer service came out well  with  account  management  (AccountBean,  LoginInfo, EmailMessage) contained in it which is desired.
\\\cline{3-4}
& & CHGNN & R4.	Blind 1 : Order Service and Inventory Service seem to be a bit overloaded.

Blind 2 : Overall the clusters are mostly well separated and map more closely to the functionality they represent.
\\\cline{3-4}
\hline
\multirow{2}{*}{Acme-Air} & \multirow{2}{*}{Java} &  CHGNN  & R3. I would prefer Cluster 2. Cluster 2 has all the functionalities well seperated.. Especially session management service, customer service.. Booking service dependencies with flight is also well captured. Comparatively, Cluster 1 didn't separate functionalities to be called as independent service
\\\cline{3-4}
& & CHGNN & R4.	Blind1 : The seeds are put in a separate cluster in most cases and the rest of the classes are all part of a single cluster. There is clearly no separation of functionality. 

Blind2 : There are clear clusters corresponding to Authentication, Flight, Booking and Customer/CustomerService. This is easily the better of the two results.
\\\cline{3-4}
\hline
\end{tabular}
\caption{With seed constraints - For all the four applications, the reviewers chose the microservices recommendations by \textit{CHGNN} over \textit{COGCN++}.}
\label{tab:outliers}
\end{table*}

\subsubsection{Qualitative studies} 

For the study, we requested each participant to compare the results from \textbf{COGCN++} to our \textbf{CHGNN} model through the json and sunburst chart images provided. We anonymized the model details in the inputs. Overall, we found that the participants had a greater agreement with the microservice recommendations produced by CHGNN than those produced by COGCN++. Table \ref{tab:outliers} captures the participants' comments for their selection for each of the applications they evaluated. The participants pointed out scope for improvements in the CHGNN recommended clusters and they had their own suggestions about how certain programs or tables could be moved to a different service. However, compared to the COGCN++ model, they recognized that CHGNN produces more acceptable and accurate clusters. Additionally, we provide our own detailed qualitative analysis of the results for one application from each of the programming paradigms - Java (OOP) and COBOL (procedural). From the feedback received, we are positive that our work helps the developers to get closer to the ideal microservices design and can substantially reduce their migration effort. This highlights the importance of factoring in resources in addition to programs for clustering and the efficacy of our heterogeneous network. 





\subsection{Authors' Comparative analysis}






\begin{figure*}
     \centering
      \begin{subfigure}[b]{0.5\textwidth}
         \centering
         \includegraphics[width=\textwidth]{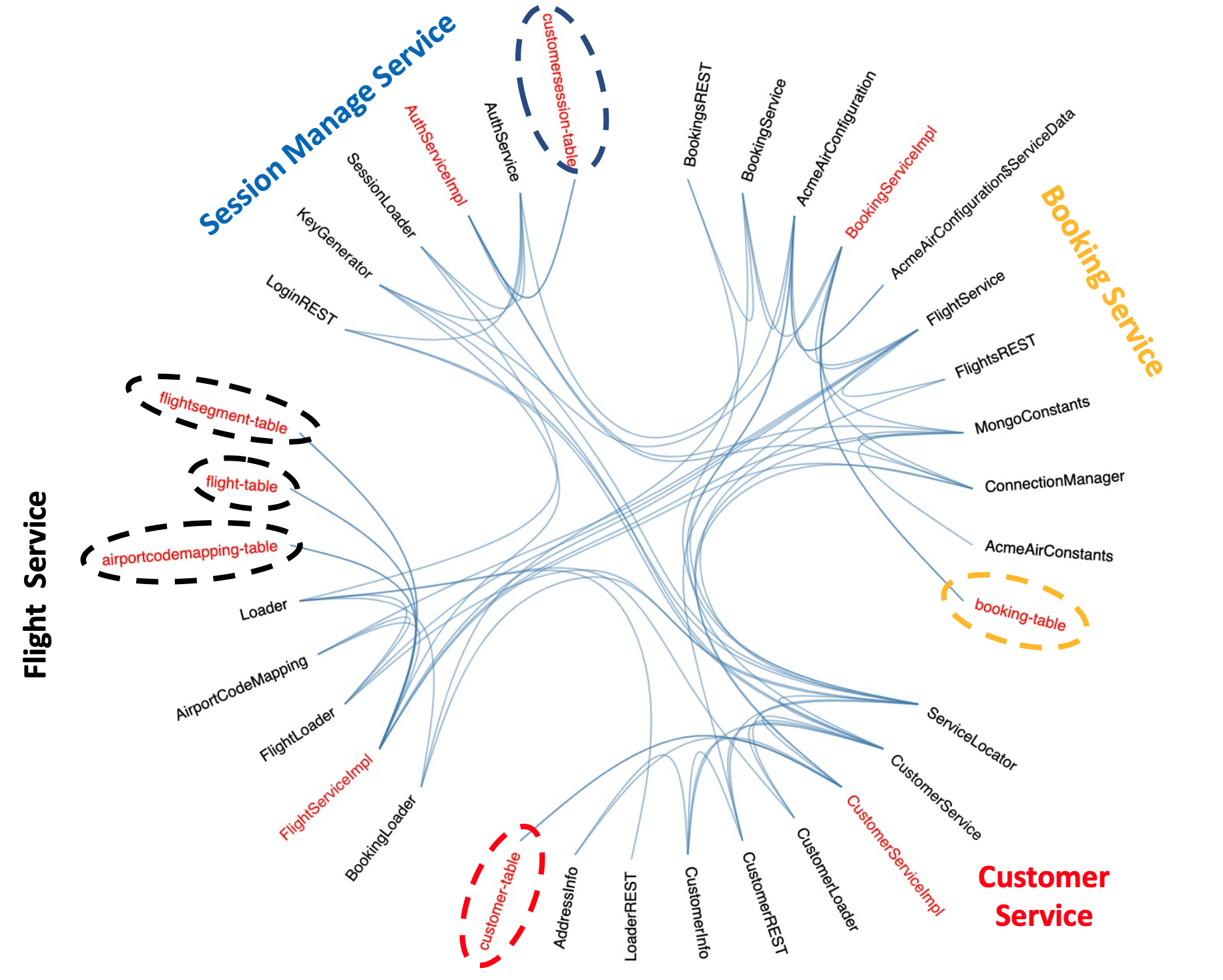}
         \caption{Results for CHGNN}
         \label{fig:acme_blind_1}
     \end{subfigure}
     \hfill
     \begin{subfigure}[b]{0.45\textwidth}
         \centering
         \includegraphics[width=\textwidth]{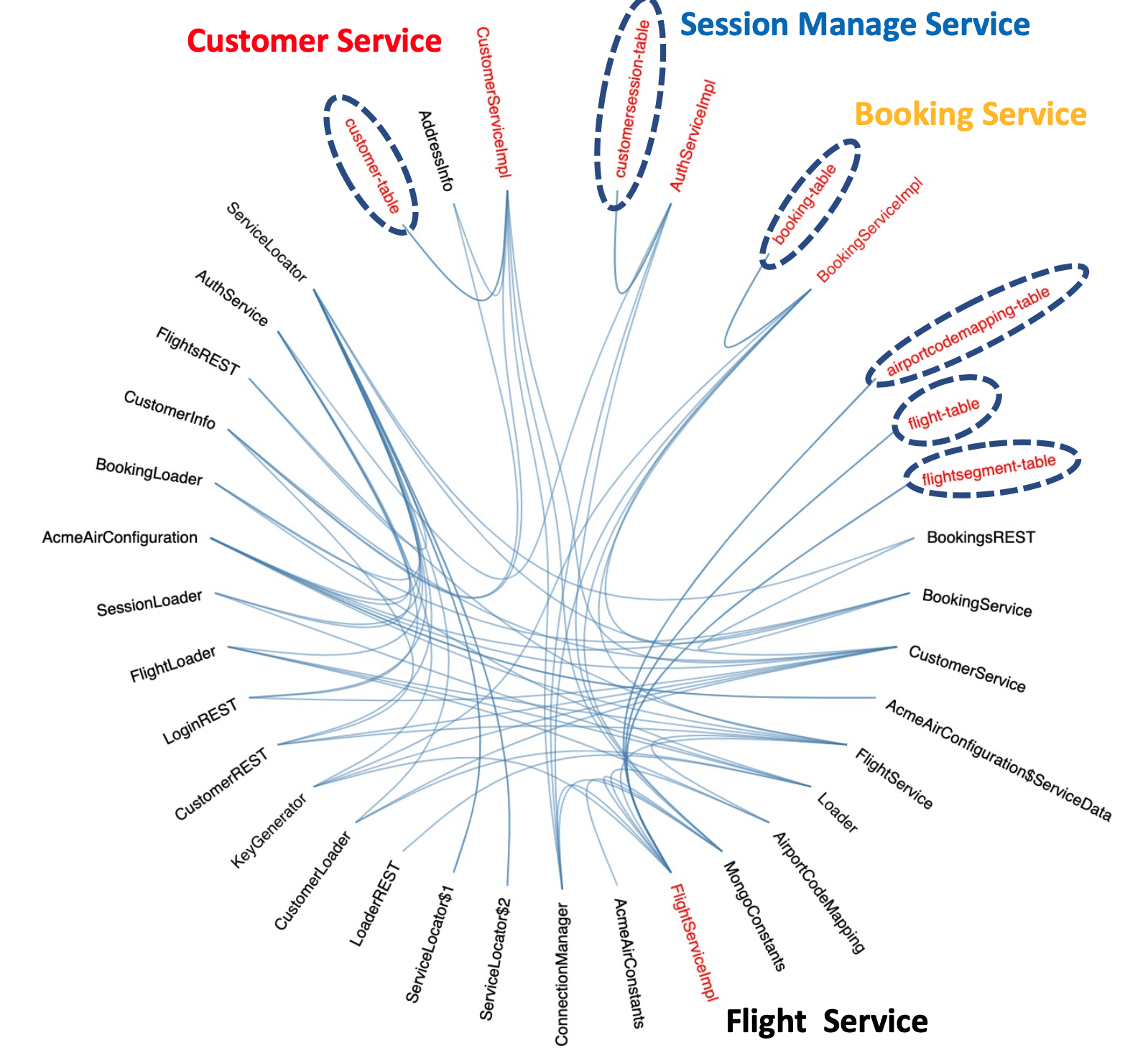}
         \caption{Results for COGCN++}
         \label{fig:acme_blind_2}
     \end{subfigure}
        \caption{Acme-Air application Sunburst charts (a) and (b) representing the microservice recommendations using CHGNN and COGCN++ models respectively.}
        \label{fig:acme-air comparison}
\end{figure*}

\begin{figure*}
     \centering
      \begin{subfigure}[b]{0.50\textwidth}
         \centering
         \includegraphics[width=\textwidth]{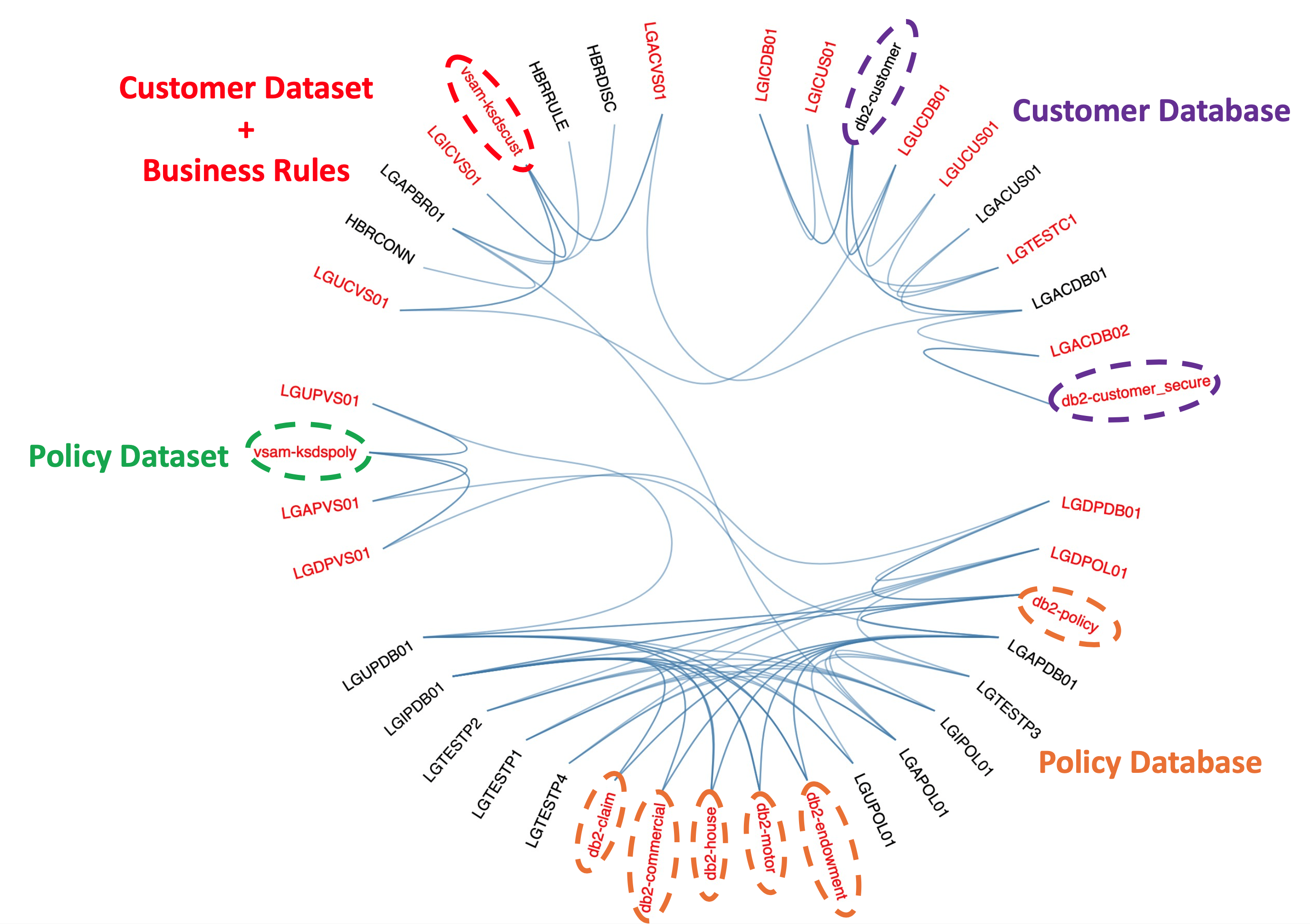}
         \caption{Results for CHGNN}
         \label{fig:genapp_blind_1}
     \end{subfigure}
     \hfill
     \begin{subfigure}[b]{0.45\textwidth}
         \centering
         \includegraphics[width=\textwidth]{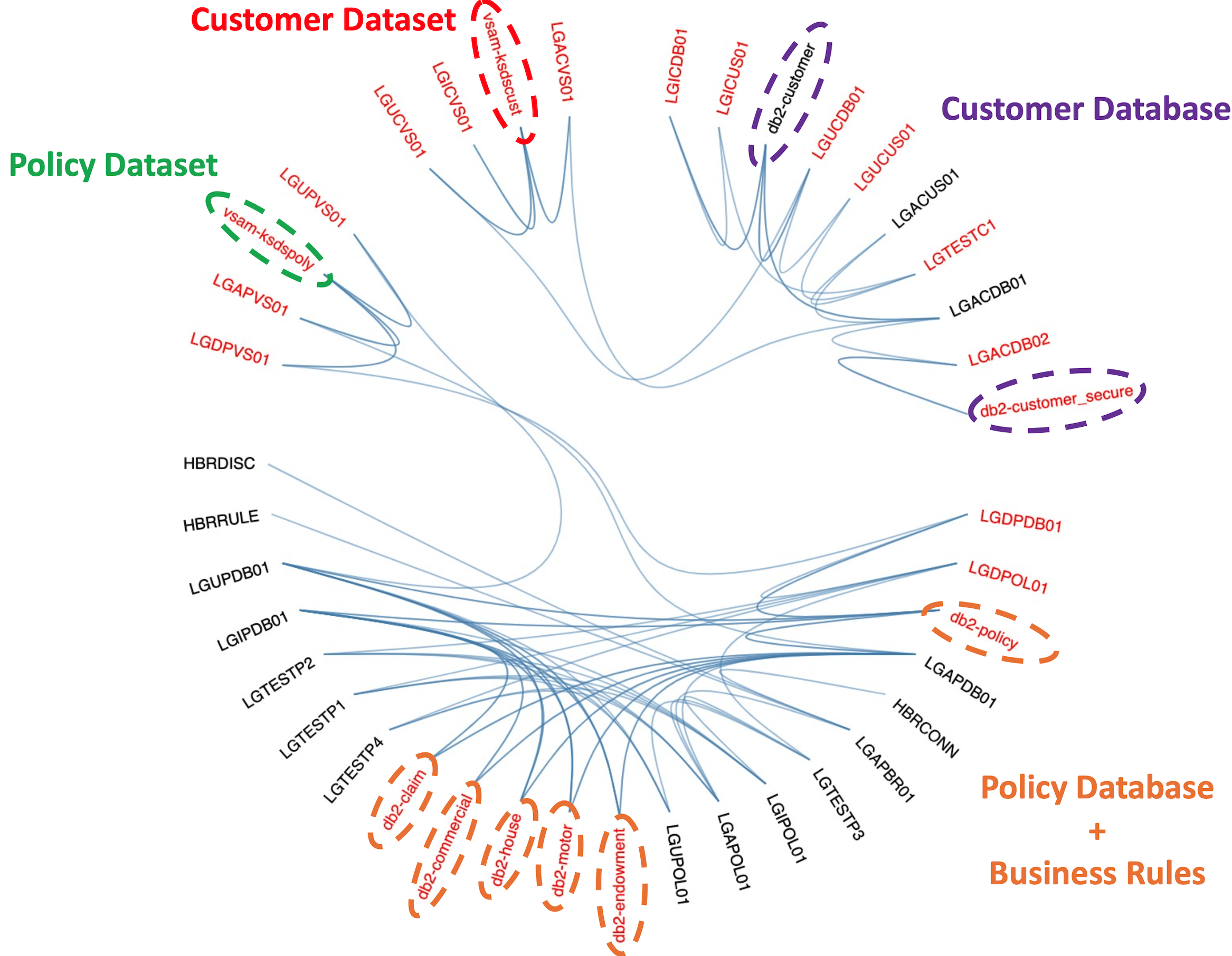}
         \caption{Results for COGCN++}
         \label{fig:genapp_blind_2}
     \end{subfigure}
\caption{GenApp application Sunburst charts (a) and (b) representing the microservice recommendations using CHGNN and COGCN++ models respectively.}
\label{fig:image2}
\end{figure*}


\subsubsection{Authors' Comparative analysis on acme-air}
Acme-Air is an application that captures key functionalities in managing an airline called "Acme Air". It contains overall 38 nodes which includes 32 java programs and 6 db tables. Figure \ref{fig:acme-air comparison} shows the output from the two models for acme-air application. We list below the differences between the two recommendations made by CHGNN and COGCN++ and provide reasons why CHGNN's output is a closer recommendation to ideal microservices.

\begin{itemize}
    \item CHGNN extracts very evenly sized clusters compared to COGCN++. This ensures that each microservice has a self-contained functionality.
    \item COGCN++ has three clusters with only two elements and one extremely large cluster. This makes the small clusters meaningless and overloads the large \textit{flight} cluster
    \item CHGNN is able to pull many relevant programs for each cluster - AddressInfo is pulled into customer service, Session Loader is pulled into session manage service etc.
    \item Due to the large \textit{flight} service in COGCN++'s output, many programs unrelated to the flight service exist. Thus, the flight service is overloaded.
      \item One drawback in CHGNN's output is that it clusters \textit{FlightService} in the booking microservice. Although \textit{FlightService} has connections with \textit{BookingServiceImpl}, it should have been aligned with \textit{FlightServiceImpl} and \textit{Flightloader} in the Flight/Airport microservice.
\end{itemize}

\subsubsection{Authors' Comparative analysis of dayTrader}

\begin{figure*} 
     \centering
      \begin{subfigure}[b]{0.7\textwidth}
         \centering
         \includegraphics[width=\textwidth]{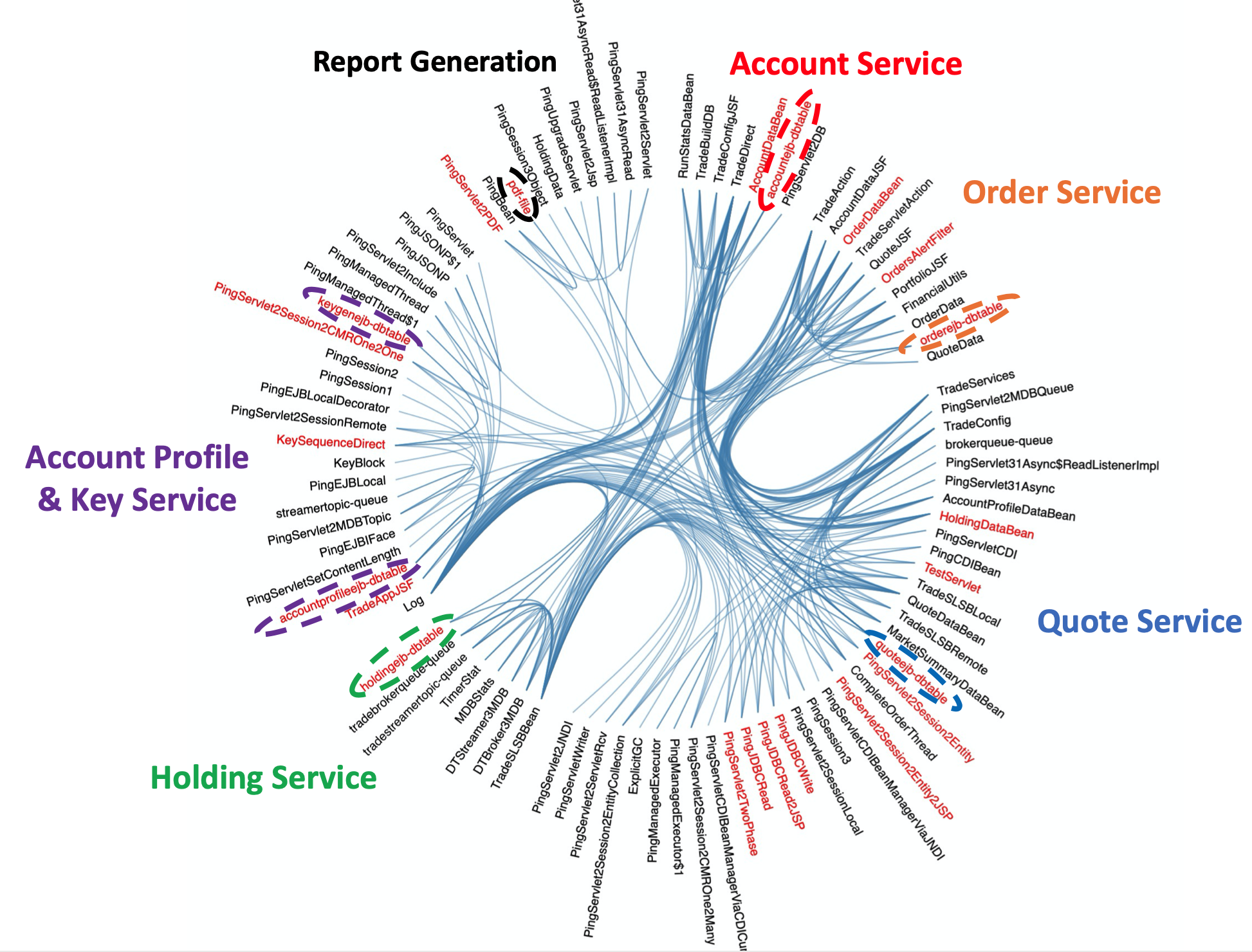}
         \caption{Results for CHGNN}
         \label{fig:dayTrader_CHGNN}
     \end{subfigure}
     \begin{subfigure}[b]{0.7\textwidth}
         \centering
         \includegraphics[width=\textwidth]{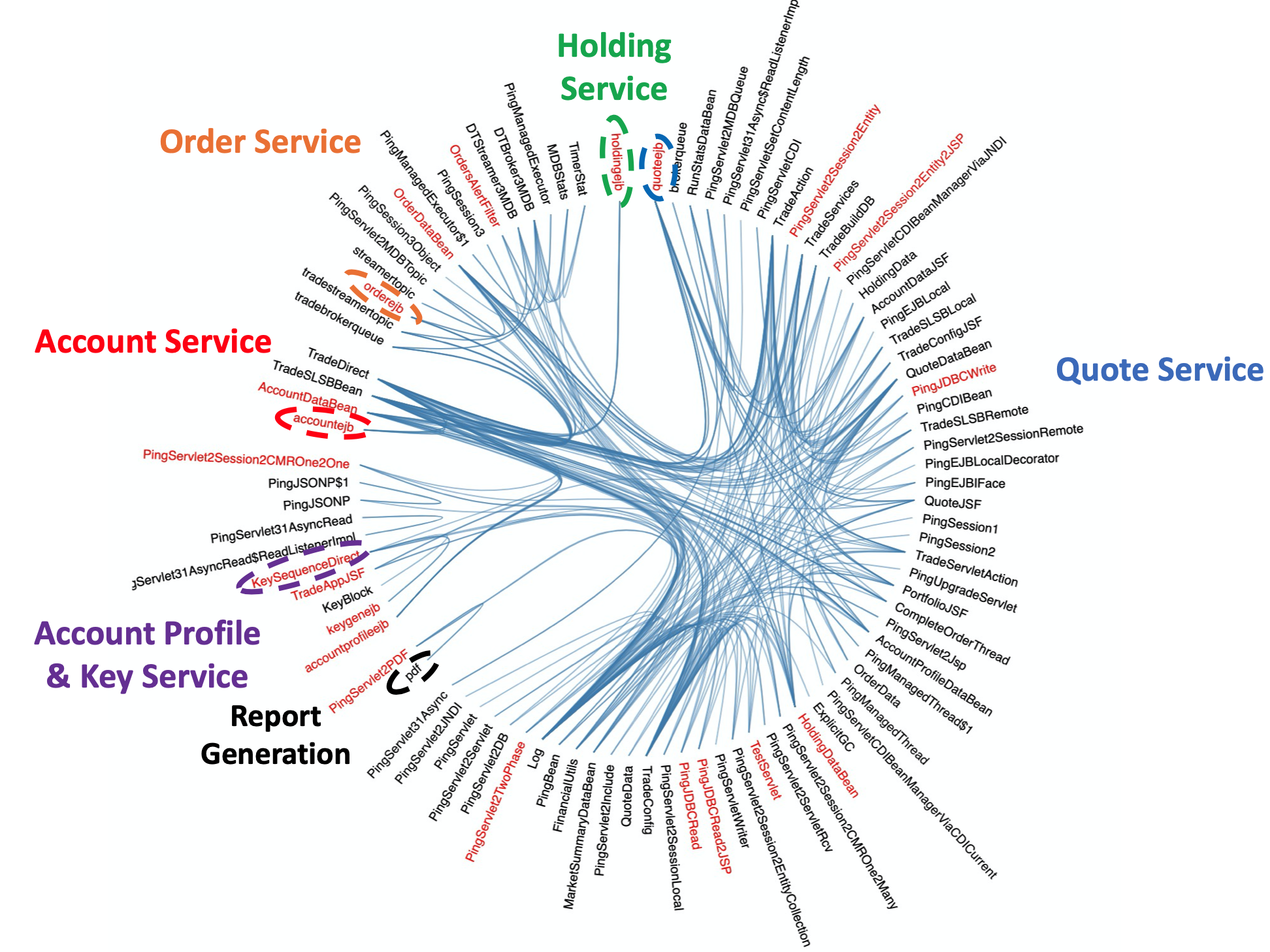}
         \caption{Results for COGCN++}
         \label{fig:dayTrader_COGCN}
     \end{subfigure}
        \caption{DayTrader application Sunburst charts (a) and (b) representing the microservice recommendations using CHGNN and COGCN++ models respectively.}
        \label{fig:dayTrader comparison}
\end{figure*}

DayTrader is an application built around the paradigm of an online stock trading system. The application allows users to login, view their portfolio, lookup stock quotes, and buy or sell stock shares. It contains a total of 122 nodes which includes 111 java programs and 11 db tables. 
Figure \ref{fig:dayTrader comparison} shows the output from the two models for the DayTrader application. We list below the differences between the two recommendations made by CHGNN and COGCN++ and provide reasons why CHGNN's output is a closer recommendation to ideal microservices.

\begin{itemize}
    \item In CHGNN's output, we see that the different functionalities are spread fairly evenly with  functional alignment.  
    \item In contrast, COGCN++'s output has an uneven distribution, with the \textit{quote} service covering more than $50\%$ of the nodes. 
    \item CHGNN's outputs have clean separations with respect to functionality - \textit{Tradesetup} which includes \textit{TradeConfig} and \textit{TradeBuild} are together along with \textit{AccountDataBean} and \textit{Account-ejb} table. The Order service which includes  \textit{OrdersFilter}, \textit{OrderData}, \textit{OrderDatabean}, \textit{Orderejb-table} are correctly clustered together. The \textit{Holding-ejb} table which is only accessed by the two key controllers -  \textit{TradeDirect} and \textit{TradeSLSBean}, has correctly been clubbed with \textit{TradeSLSBean}. While its other dependency TradeDirect is assigned with its key dependent Trade setup programs. Similarly report generation which is handled by \textit{PingServlet2pdf} is clubbed separately. Thus, overall, our approach manages to get closer to ideal microservices that 1) Have programs that are functionally coherent and 2) Have key dependent files and resources that closely interact with these programs.
    \item In COGCN++'s output, we do not see as much functional coherence amongst the programs in a cluster. 
    
\end{itemize}

\subsubsection{Authors' Comparative analysis on GenApp}
Figure \ref{fig:image2} showcases the resulting clusters for GenApp by CHGNN and COGCN++ models. Functionally, GenApp is an application that creates insurance policies and processes customer-claims. Hence it has two types of databases and datasets - the customer database \& customer dataset, as well as a group of policy databases (house, motor, vehicle etc.) and a policy dataset. This entire codebase is broken into two main logical groups (containing code and data) - the $1^{st}$ deals with customer acquisition and the $2^{nd}$ deals with policies bought and claimed by customers. It is implemented as a monolithic COBOL codebase that contains $30$ program nodes and $10$ resource nodes. Hence, on the whole, resource nodes contribute to $25\%$ of all available nodes in the software. After running COGCN++ and CHGNN on Genapp, we carefully analyse both outputs and observe the following points.

\begin{itemize}
    \item We notice that both COGCN++ and CHGNN are successfully able to identify and isolate the \textit{Customer Dataset}, \textit{Policy Dataset}, \textit{Policy Database} and \textit{Customer Database} services.
    \item However, unlike CHGNN that clubs the \textit{business rules} functionality with \textit{Customer Dataset} - COGCN++ chooses to club it with \textit{Policy Database}. As a result, COGCN++ creates a huge cluster that dominates the other clusters with respect to the number of artifacts. This size imbalance has many practical implications on the ground - as detailed by \textit{R2} in Table \ref{tab:outliers}. 
    
\end{itemize}

\subsection{Without Seed Constraints}

\begin{figure*}
     \centering
      \begin{subfigure}[b]{0.45\textwidth}
         \centering
         \includegraphics[width=\textwidth]{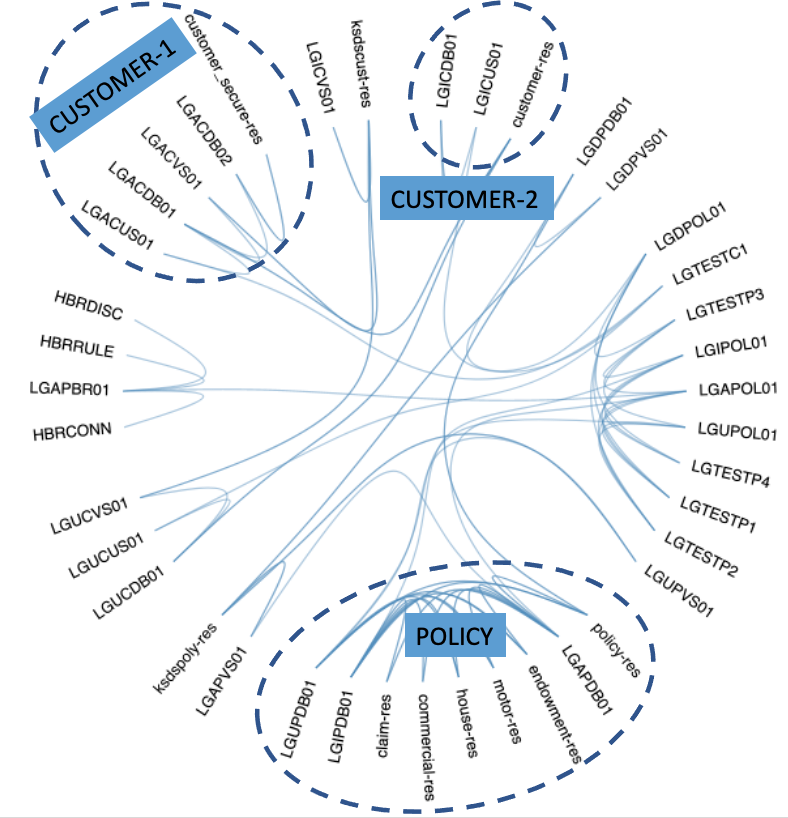}
         \caption{Results for CHGNN}
     \end{subfigure}
     \hfill
     \begin{subfigure}[b]{0.5\textwidth}
         \centering
         \includegraphics[width=\textwidth]{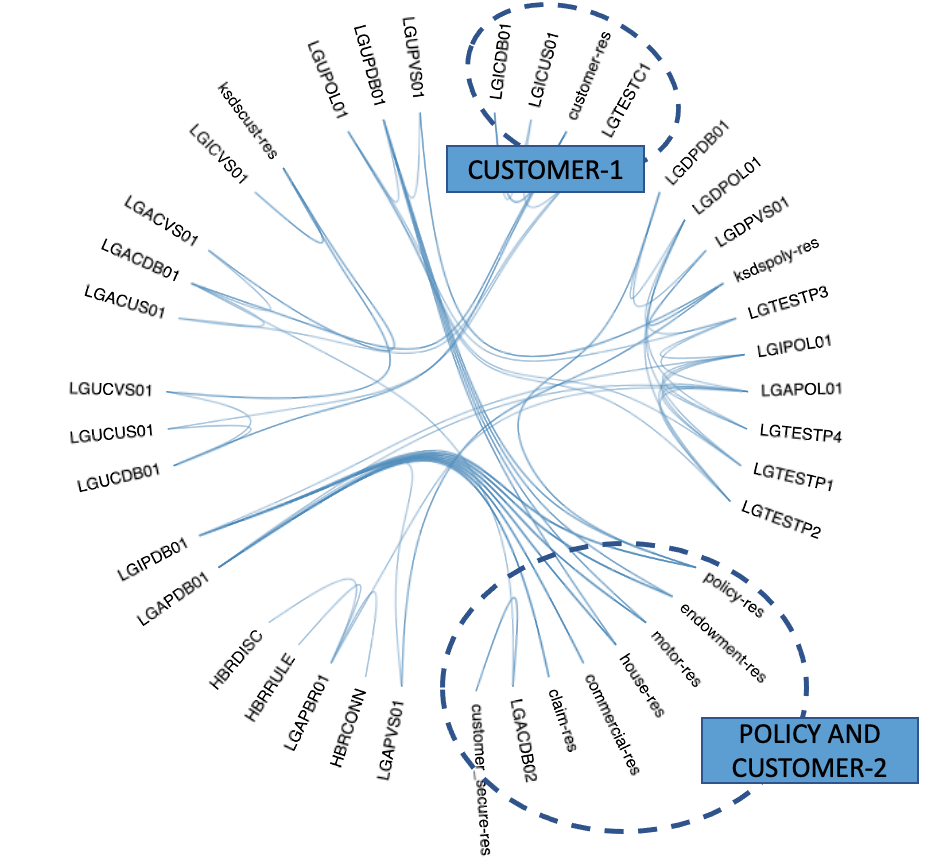}
         \caption{Results for COGCN}
     \end{subfigure}
\caption{GenApp application Sunburst charts (a) and (b) representing the microservice recommendations using CHGNN and COGCN models respectively.}
\label{fig:image5}
\end{figure*}

We also experiment with CHGNN by \textit{not} specifying the seed constraints and note down the results in this section. One of the drawbacks of using a GNN without seed constraints, is that we cannot guarantee a clean separation of the application's distinct functionalities. As a result, there are instances where the output clusters have more/less than one functionality. Hence, the clusters obtained from the GNN without seed constraints may not be cleanly differentiable. Note that in the absence of SME inputs for $K$ (and seeds), we choose that value of $K$ which maximises modularity of the communities. For example, in Figure \ref{fig:image5}, $K=9$ for Genapp. Although this maximises the modularity metric - there is no clear \textbf{functional role} for each cluster. This is the main motivation behind incorporating deep domain knowledge with the help of seeds. 

\textbf{Note} : As we do not use seed constraints in the following results, we use the original vanilla COGCN (and not \textbf{COGCN++}).

We show the reviewer's qualitative analysis of these clusters in Table \ref{tab:outliers2}. In this analysis as well, all reviewers agree that CHGNN is a better microservice recommender than COGCN. We additionally show the clusters generated for Genapp in Figure \ref{fig:image5}. As depicted in Figure \ref{fig:image5}, we see that CHGNN has better formed clusters than COGCN, as it is able to cleanly separate between the customer and policy functionality. But COGCN ends up mixing some functionalities of customer and policy together. However, both results are still inferior to the results in Figure. \ref{fig:image2}.




\begin{table*}[t]
\small
\centering
\begin{tabular}{|p{0.1\textwidth}|p{0.1\textwidth}|p{0.1\textwidth}|p{0.6\textwidth}|}\hline
\textbf{Application} & \textbf{Language} & \textbf{Reviewer Selection} & \textbf{Reviewer Reasons} \\
\hline
\multirow{2}{*}{GenApp} & \multirow{2}{*}{COBOL} &  \multirow{2}{*}{CHGNN}  &  R1. I have analysed both sets of outputs on 3 aspects. A) Number of datasets or tables correctly grouped in the cluster (more the better), B) Number of programs in the cluster that are incorrectly placed (less the better) and C) Number of clusters that do not have a well defined function (less the better). On all 3 aspects, I find that the blind1.json is a better result.	
\\\cline{3-4}
& & \multirow{2}{*}{CHGNN}  &  R2.  In Blind-1, my biggest problem is that TESTC1 comes with a Policy cluster. In Blind-2, my main problems are 1) LGAPVS01 is grouped wrong. 2) LGACDB02 and Customer-Secure table are grouped wrong 3) All tables are grouped together and LGAPVS01 is thrown in as a bonus. Relatively speaking, Blind-1 is far better.
\\\cline{3-4}
\hline
\multirow{2}{*}{Daytrader} & \multirow{2}{*}{Java} &  CHGNN  & R3. Keysequence, KeysequenceDirect, keygen table are all together. TradeSLSBean and TradeDirect which are common in functionality are put together. Blind2 has a very large cluster with tables like keygen packed together which looks wrong. But still Blind2 also should be improved to group TradeConfig and TradeAction
\\\cline{3-4}
& & CHGNN & R4.	Overall I think it is a good separation of functionality, I could associate a name with each cluster to some extent. Only a few things bothering me: most db tables end up in the same cluster as TradeDirect, but since TradeSLSBean is also present here, it probably makes sense. Also the ping classes assigned to the cluster with Order and Holding data seem a bit random.
\\\cline{3-4}
\hline
\multirow{2}{*}{PBW} & \multirow{2}{*}{Java} &  CHGNN  & R3. Reason : supplier service came out well with its db contained within it. Inventory, Shopping and backorder which are closely connected are clubbed together with inventory and backorder table together. Also in Blind2, there is no good explanation for catalogmanager and emailmessage
\\\cline{3-4}
& & CHGNN & R4.	In Blind 1, there are clear Supplier, Inventory/Backorder and Customer clusters with the only issue being customer-db moved to a different cluster. Blind 2 on the other hand has no clear separation of clusters. Inventory/Backorder and Shopping seem mixed up.\\\cline{3-4}
\hline
\multirow{2}{*}{Acme-Air} & \multirow{2}{*}{Java} &  CHGNN  & R3. Reason : Customer service came out as a well separated cluster. I find that unlike Blind2 it didn’t mix the booking with authentication service. At a first glance, I was confused why FlightService abstract class is separated from FlightServiceImpl and kept with BookingService. But BookingService seem to have dependency only to the only FlightService implemented method getFlightByFlightId.
\\\cline{3-4}
& & CHGNN & R4.	Flight and Customer clusters look more complete in Blind 1 and other clusters look better overall. In Blind 2, the customer table seems misplaced and the bottom left and bottom right clusters seem to be a bit overloaded.\\\cline{3-4}
\hline
\end{tabular}
\caption{Without seed constraints - For all the four applications, the reviewers chose the microservices recommendations by \textit{CHGNN} over \textit{COGCN}.}
\label{tab:outliers2}
\end{table*}

\begin{table*}[ht]
\centering
\resizebox{0.9\linewidth}{!} {
\begin{tabular}{l|lllll|llllll}
\hline
& \multicolumn{5}{c|}{\textbf{ACME} (Type:Airline App, Lang:Java)} &  & \multicolumn{5}{c}{\textbf{DayTrader} (Type:Trading App, Lang:Java)} \\
    & \multicolumn{5}{c|}{(K=4, \#Class=30, \#Resource=6)} &  & \multicolumn{5}{c}{(K=6, \#Class=111, \#Resource=11)} \\
    & Mod ({$\uparrow$})   & NED ({$\uparrow$})   & S-Mod ({$\uparrow$})  & Coverage ({$\uparrow$}) & $Metric_{Sum}$   &  & Mod ({$\uparrow$})   & NED ({$\uparrow$})   & S-Mod ({$\uparrow$})  & Coverage ({$\uparrow$}) & $Metric_{Sum}$ \\ \hline

COGCN++  &0.106   & 0.044  & \textbf{0.379} & \textbf{0.8}  & 1.329 &  & 0.135	& 0.226 & \textbf{0.229} & \textbf{0.566} & 1.156  \\
HetGCNConv &\textbf{0.247} & 0.635  & 0.233  & 0.597 &  1.712 &  & 0.073	& \textbf{0.566}	& 0.133	& 0.324 & 1.096 \\
CHGNN-EL & 0.237 & \textbf{0.738} & 0.207 & 0.592   &  1.775 &  & 0.168	& 0.549	& 0.173	& 0.421 & 1.310 \\
CHGNN & 0.246  & \textbf{0.738} & 0.214 & 0.586     & \textbf{1.784} &  & \textbf{0.175}	& 0.549  & 0.177 & 0.435 & \textbf{1.336} \\ \hline

& \multicolumn{5}{c|}{\textbf{PBW} (Type:Plant Store, Lang:Java)} &  & \multicolumn{5}{c}{\textbf{Genapp} (Type:Insurance App, Lang:Cobol)} \\
& \multicolumn{5}{c|}{(K=5, \#Class=30, \#Resource=6)} &  & \multicolumn{5}{c}{(K=4, \#Program=30, \#Resource=10)} \\
& Mod ({$\uparrow$})   & NED ({$\uparrow$})   & S-Mod ({$\uparrow$})  & Coverage ({$\uparrow$})  & $Metric_{Sum}$ & & Mod ({$\uparrow$})   & NED ({$\uparrow$})   & S-Mod ({$\uparrow$})  & Coverage ({$\uparrow$}) & $Metric_{Sum}$\\ \hline
COGCN++                    & \textbf{0.209}  & 0.433 & \textbf{0.351}  & \textbf{0.59}      &  1.583 & &  0.424  & 0.25   & \textbf{0.277}   & \textbf{0.919} & 1.870 \\
HetGCNConv             & 0.161 & 0.649  & 0.274  & 0.493  & 1.577  &&  0.378  & \textbf{0.704} & 0.222   & 0.711 & \textbf{2.016} \\
CHGNN-EL & 0.195	& \textbf{0.858} & 0.272 & 0.479      &  \textbf{1.805} & &  \textbf{0.443}	& 0.444  & 0.266 & 0.856 & 2.010 \\
CHGNN    & 0.194 & 0.830 & 0.272 & 0.466 & 1.762 & &  0.440	& 0.445  & 0.269 & 0.846 & 2.000 \\

\bottomrule
\end{tabular} 
}
\caption{Performance of partitioning Monolith to Microservices (results averaged over $30$ runs)}
\label{tab:metrics2}
\end{table*}

\section{Quantitative Analysis}

\subsection{Quantitative Metric Implementation}

We leverage many out-of-the-box functionalities from the NetworkX library\footnote{https://networkx.org/}. For \textit{modularity} (Mod), we leverage the moduarity function in NetworkX\footnote{https://networkx.org/documentation/stable/reference/algorithms/generated/networkx.algorithms.community.quality.modularity.html}. For \textit{coverage}, we leverage the coverage function in NetworkX\footnote{https://networkx.org/documentation/stable/reference/algorithms/generated/networkx.algorithms.community.quality.coverage.html}. For S-Mod and NED, we implement the metrics from scratch using NetworkX graph objects and methods.

\subsection{Trade-off between Metrics}
Due to inherent trade-offs between the metrics, we did not study each metric independently in the main paper. This is because it is possible to choose specific strategies that inflate certain metrics at the expense of underperforming other metrics. For example, by creating few \textit{boulder} (very large) clusters and many \textit{dust} (very small) clusters, we can increase intra-cluster edges and decrease inter-cluster edges. This would improve S-Mod and Coverage considerably. However, this would also score poorly on NED - which penalises for size imbalance. Hence, instead of relying on any one metric, we calculated $Metric_{Sum}$ (sum of all metrics) and ranked our approaches accordingly.

\subsection{Finer Analysis}
In this section, we further qualify our quantitative metric results as depicted in Table \ref{tab:metrics2}. In what follows, we detail two clear trends that we have observed. Some of the explanations below will involve referring to Figure. $2$ in the main paper.

(i) COGCN++ has the highest S-Mod and Coverage values but simultaneously has the lowest NED score in all four apps. This is because unlike CHGNN that creates evenly sized clusters, COGCN++ creates a few \textit{boulder} clusters like the \textit{Order} cluster (having $19$ artifacts) and many \textit{dust} clusters like \textit{Supplier}, \textit{BackOrder} and \textit{Customer} (having atmost $3$ artifacts). This intuitively results in more \textit{intra-cluster} edges and less \textit{inter-cluster} edges - thus inflating Coverage and S-Mod. However, this size imbalance is penalized by NED. This also validates our corresponding qualitative observations in PBW where we pointed out that COGCN++ outputs clusters having lesser meaning and overloaded functionality.

(ii) In regard to the Mod metric, there is no obvious winner. However, we observe that (1) CHGNN or CHGNN-EL consistently feature in the top two results across all four apps and (2) COGCN++ places amongst the bottom two in three of the four apps. For PBW, although COGCN++ outperforms on Mod, many of the predicted clusters are overloaded with functionality and have lesser meaning (refer Figure 2 in main paper). Thus a quantitative out performance on metrics does not necessarily translate to better functionally aligned clusters. 

\section{Limitations}
Like we discussed in the conclusion section, the current method works on the program and resource level but the decomposition task can be studied at a more granular level from \textit{programs} to \textit{functions} and \textit{tables} to \textit{columns}. Also, we had to limit our study to only four medium sized applications as it takes ~3-4 weeks for a software developer to understand the implementation structure of the application and provide feedback for the generated recommendations. However, based on the positive feedback from the qualitative analysis, we believe that our approach can significantly reduce developers' effort in finalizing the ideal microservices design for the  migration activity. 

\end{document}